\documentclass[fleqn,twoside]{article}
\usepackage{espcrc2}
\usepackage{graphicx}
\usepackage{epsfig}

\usepackage[figuresright]{rotating}

\newcommand{\AmS}{{\protect\the\textfont2
  A\kern-.1667em\lower.5ex\hbox{M}\kern-.125emS}}

\title{The Chroma Software System for Lattice QCD}

\author{  Robert G. Edwards\address[MCSD]{Thomas Jefferson National Accelerator Facility, Newport News, VA 23606, USA} {\em SciDAC and LHPC Collaboration},
B\'alint Jo\'o\address[MCSD]{School of Physics, University of Edinburgh, Edinburgh, Scotland, UK, EH9 3JZ} {\em UKQCD Collaboration}
}

\begin{document}

\begin{abstract}
We describe aspects of the Chroma software system for lattice QCD calculations.
Chroma is an open source C++ based software system developed using the software infrastructure of the US SciDAC initiative.
% The system uses SciDAC packages such as QDP++ for data parallel operations, QMP for parallell communications and 
%QIO for binary I/O. XML I/O is provided by wrappers areount the open source libxml2 library. 
Chroma interfaces with output from the BAGEL assembly generator for optimised lattice fermion kernels on some architectures. It can be run on workstations, clusters and the QCDOC supercomputer. 
%The majority of the development effort was carried out at the JLab with strong collaboration from UKQCD. 
% As a case study we briefly describe the use of Crhoma in our investigations fo the Overlp Fermion Matrix.
\end{abstract}

\maketitle
\section{INTRODUCTION}
We present the Chroma software system \cite{Chroma} for lattice QCD (LQCD) calculations. Chroma aims to provide a computational LQCD toolbox which is flexible, portable and efficient on a wide range of architectures from desktop workstations to clusters, commercial machines and new architectures such as the QCDOC \cite{QCDOC}

Development on Chroma started at the JLab\footnote{in full:
Thomas Jefferson National Accelerator Facility} for the U.S. lattice
community, in particular the LHPC collaboration \cite{LHPC}, using
software from the U.S. SciDAC initiative \cite{SciDAC}. This effort
has been joined by the UKQCD collaboration \cite{UKQCD} who have been
contributing to the effort on all levels.
% The result of this collaboration has
% not only produced useful software but also highlights the benefits of
% large-scale international collaboration and the open source
% development process.

%The Chroma suite is a high level package, concentrating on the physical 
%problems to be solved.
To achieve the goals of flexibility, portability and efficiency, Chroma relies on several layers of SciDAC and UKQCD software.

%%% You can change the following value of 25cm if you wish, to make your
%%% columns wider or narrower.  For example, you may wish to generate some
%%% narrow columns of text in order to "flow" around a figure
\subsection{SciDAC Software Hierarchy}
Toward the end of 2000, the U.S. Lattice community embarked on an ambitious
project through the U.S. SciDAC initiative to standardise a set of software
components in order to allow the effective exploitation of computing resources
for LQCD. The following levels of software infrastructure were defined:
\begin{description}
\item{\bf QCD Message Passing (QMP): \ } provides a message
passing API  tailored to LQCD calculations. QMP was designed to take
advantage of the specialised communication hardware of emerging
architectures, such as the Serial Communications Unit (SCU) of the
QCDOC, or the capabilities of Myrinet Network devices.
\item{\bf Level 1 -- QCD Linear Algebra (QLA): \ } defines operations to be performed at each site of the lattice. Primitives include $SU(3)$
matrix--matrix and matrix--colour vector multiplications.
\item{\bf Level 2 -- QCD Data Parallel (QDP): \ }  provides lattice-wide operations such as basic linear algebra for lattice-wide fields.
\item{\bf Level 3 -- Special optimised software: \ } will provide portable interfaces to highly optimised and machine-dependent pieces of code such as assembly-coded Dslash operators.
\end{description}
A recent addition to this hierarchy is the {\bf QCD I/O (QIO)}
sublayer which provides a record oriented I/O API. XML metadata
and binary data can be packaged as separate records of a single file
for transmission to future Web/Grid services.

%
%We illustrate the hierarchy in Fig. 1. Software for QMP, QLA, QDP and QIO are available from the SciDAC web site \cite{SciDAC}. Chroma currently relies on the QMP layer, and a C++ implementation of the QDP layer called QDP++.

%\begin{figure}[htb]
%\epsfxsize=75mm
%\epsffile{scidac_layers.eps}
%\caption{The SciDAC Software Hierarchy}
%\label{fig:MWfig}
%\end{figure}

%%% You can change the following value of 25cm if you wish, to make your
%%% columns wider or narrower.  For example, you may wish to generate some
%%% narrow columns of text in order to "flow" around a figure
Level 3 of the SciDAC hierarchy is not yet as mature as the
other levels, hence Chroma uses optimised Dslash operators from a
variety of sources, such as the Pentium 4 SSE assembler code developed at the JLab  \cite{SSEDslash}, or the output of the BAGEL \cite{BAGEL} generator. 

%The BAGEL assembly generator is a third party, non-SciDAC, package
%which models the memory access, functional units and pipelines of the
%target processor and cache system taking into account their latencies
%and execution times. By implementing agressive prefetching strategies
%it produces extremely fast code for RISC architectures.

%For RISC architectures such as the QCDOC and POWER based systems we
%have interfaced to code produced by the BAGEL assembly generator
%\cite{BAGEL} which produces highly optimised assembler kernels for
%RISC microprocessors by modelling the CPU pipelines and taking into
%account the latencies and execution times of the CPU functional units.
%BAGEL also implements aggressive prefetching schemes resulting in
%extremely fast code. While BAGEL is not SciDAC software, it is in fact
%a very versatile tool that can be used to generate operators for
%SciDAC's level 3. The source code for BAGEL will be available for
%download in the future \cite{BAGEL}.

%%% You can change the following value of 25cm if you wish, to make your
%%% columns wider or narrower.  For example, you may wish to generate some
%%% narrow columns of text in order to "flow" around a figure
\section{QDP++}
QDP++ is a C++ implementation of the QDP level of the SciDAC hierarchy, which 
Chroma uses as its foundation. QDP++ defines lattice-wide types and allows
lattice-wide expressions. It has been integrated with the QIO framework and with modules to provide XML based I/O (e.g. for parameter reading, or processing QCDML markup \cite{ILDG}).

\subsection{Lattice--Wide Types}
%%% You can change the following value of 25cm if you wish, to make your
%%% columns wider or narrower.  For example, you may wish to generate some
%%% narrow columns of text in order to "flow" around a figure
QDP++ models the tensor product structure of LQCD objects through
a series of nested templates. E.g: the indices of lattice fermion fields can follow the structure:
\begin{eqnarray*}
\mbox{\em sites} \otimes \mbox{\em spins} \otimes \mbox{\em colours}  \otimes \mbox{\em complex numbers}
\end{eqnarray*}
QDP++ would model this through the following C++ templated type:
{\normalsize \begin{verbatim}
OLattice< PSpinVector< PColorVector< 
 RComplex < REAL >, Nc >, Ns > >
\end{verbatim}}
where {\tt REAL} is the type of the complex components, defined as either {\tt float} or {\tt double}. {\tt Nc} and 
{\tt Ns } are the numbers of spin and colour components defined during configuration.

QDP++ operates on such types by recursing down the tower of templates. To
multiply the above fermion type by a scalar, one would first loop through the
indices of the {\tt OLattice} type and for each one, call the
multiply operation for the {\tt PSpinVector} type and so on.

The order of the templates is not fixed in principle and could be
permuted in order to take advantage of different parallel
architectures. The templated types are aliased to fixed type names
such as {\tt LatticeFermion} that are used by higher-level codes.

\subsection{Template Expressions, PETE}
%%% You can change the following value of 25cm if you wish, to make your
%%% columns wider or narrower.  For example, you may wish to generate some
%%% narrow columns of text in order to "flow" around a figure
QDP++ provides lattice-wide arithmetic expressions by overloading operators. The AXPY operation 
\begin{equation}
 z \leftarrow \alpha x + y \label{eq:AXPY}
\end{equation}
for lattice fermion fields $x, y, z$ and scalar $\alpha$ is coded as:
\begin{equation} 
  \mbox{\tt z = a*x + y} .
\end{equation}

\begin{figure}[htb]
\epsfxsize=75mm
\epsffile{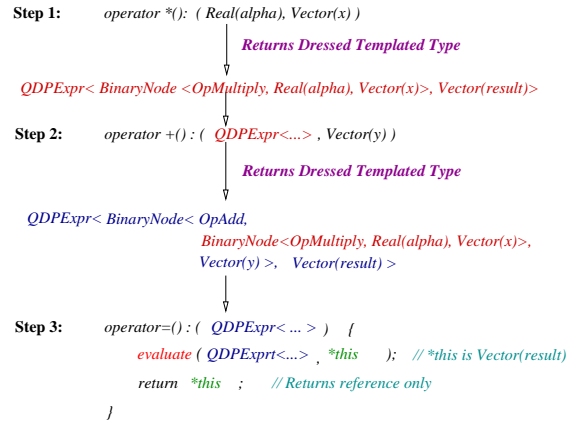}
\caption{Expression Transformation}
\label{fig:PETEExp}
\end{figure}

To eliminate temporaries in expressions, QDP++ uses the Portable
Expression Template Engine (PETE)\footnote{PETE is a stand alone component of the POOMA library} \cite{PETE}.  The C++ binary operators are
redefined to transform the expression into an {\bf expression template
type}. On assignment, storage for the result and the expression
template, which includes references to the operands, are all given as
{\em arguments} to an {\em evaluate()} function. {\bf This is done at
{\em compile time} through the template instantiation mechanism}. The 
process for the AXPY operation of Eq. (\ref{eq:AXPY})  is shown in Fig. \ref{fig:PETEExp}
\vspace{-0.5cm}

% \subsection{QDP++ Optimisation}
%%% You can change the following value of 25cm if you wish, to make your
%%% columns wider or narrower.  For example, you may wish to generate some
%%% narrow columns of text in order to "flow" around a figure
Templated {\em evaluate()} functions can be {\bf specialised} for specific expression templates, and the specialised functions can  call optimised
subroutines, e.g. fast AXPY like functions. 

%This approach is in
%some sense of limited use, since only a few hand--picked expressions
%are optimised. However, in practice, only a small number of routines
%(e.g. Level 1 BLAS) are of greatest utility and of most common use. To this
%end we have optimised real and complex AXPY routines, inner products 
%and vector norm operations.
It may be possible to optimise general expressions even further  by applying
PETE techniques at all levels of our template hierarchy rather than
just at the {\tt OLattice} level as is done now.

%%% You can change the following value of 25cm if you wish, to make your
%%% columns wider or narrower.  For example, you may wish to generate some
%%% narrow columns of text in order to "flow" around a figure
\section{CHROMA}
Chroma developed to serve the needs of the LHPC and
UKQCD collaborations, which have included {\em spectroscopy,
decay constant, nucleon form factor} and {\em structure function moment}
calculations. The authors are also interested in investigating {\em chiral fermion actions}. Hence the code contains {\em Wilson, Domain Wall} and {\em Overlap
fermion operators} and {\em numerous inverters}. Code also exists to compute
{\em hadronic 2-point} and {\em 3-point} correlation functions.  Furthermore,
UKQCD researchers have written an {\em ASqTAD fermion
operator} in the staggered library.

\subsection{Chroma and Efficiency}
By optimising QDP++ as discussed previously and using high performance kernels in our linear operators we have achieved high performance on workstations, clusters and the QCDOC. 

\begin{figure}[htb]
\epsfxsize=70mm
\epsffile{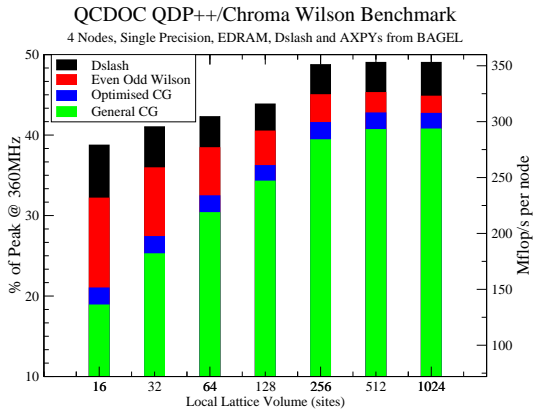}
\caption{Performance on the QCDOC}
\label{fig:ChromaPerf}
\end{figure}

Fig. \ref{fig:ChromaPerf} shows single-precision performance
measurements with an even-odd preconditioned Wilson Dirac operator on
4 nodes of a QCDOC using the assembler Dslash from BAGEL. We show the
performance of the Dslash, the Dirac operator, a naively implemented
Conjugate Gradients (CG) solver and an optimised CG solver, which makes
direct calls to the Dslash routine and fuses the application of
the Dirac Matrix with the computation of the norm of the result where
possible.

The Dslash routine (from BAGEL) was designed to amortise
communications latency at a local lattice size of 256 sites. Scaling
from this point to larger volumes can be seen in
Fig. \ref{fig:ChromaPerf}. The cost of the additional vector
operations needed to combine the Dslash applications into the Dirac
operator appears as the difference between the red and black bars.  The
difference between the general CG and the optimised one is about 2-3\%
of peak over the whole range of volumes, showing that the overhead
from QDP++ expressions is very small.
\vspace{-3pt}
\vspace{-30pt}

\section{CONCLUSIONS}
%%% You can change the following value of 25cm if you wish, to make your
%%% columns wider or narrower.  For example, you may wish to generate some
%%% narrow columns of text in order to "flow" around a figure
We have discussed the SciDAC software hierarchy and elements of the
QDP++ and Chroma software suites for lattice QCD and have shown how
the latter may be optimised in order to achieve high efficiency
through template expressions and by interfacing cleanly with third
party high performance libraries.
%We have also 
%illustrated the ease of implementing new solver algorithms. 

Due to lack of space we have had to gloss over several other nice
features of Chroma and QDP++ such as the many applications already
existing in the code base and our build system which uses the GNU
Autotools, allowing for easy configuration and building. These 
details are deferred to a forthcoming publication.

%We believe Chroma has a bright future. Already the LHPC and UKQCD
%collaborations are using it for production calculations on a variety
%of clusters and it has been ported and partially optimised for the
%QCDOC.

Our future plans include the implementation of gauge configuration
generating algorithms for quenched and dynamical fermions. At the time
of writing, the basic class framework for this is already in place.

QMP, QDP++ and Chroma are {\em open source software and are freely available
through anonymous CVS,} details of which can be found in \cite{Chroma}.

\section{ACKNOWLEDGEMENTS}
R.~G.~Edwards is supported under DOE contract DE-AC05-84ER40150 under
which the Southeastern Universities Research Association (SURA)
operates the Thomas Jefferson National Accelerator Facility. B.~Jo\'o gratefully acknowledges funding under
PPARC Grant PPA/G/0/2002/00465.  We thank P.~A.~Boyle for assistance with the BAGEL assembly generator.

\end{document}